\documentclass[amssymb,amsmath,pra,floatfix,twocolumn]{revtex4}
\usepackage{algorithmic}
\usepackage{algorithm}
\usepackage{url}
\usepackage{graphicx}     

\newtheorem{observation}{Observation}
\newtheorem{definition}{Definition}
\newtheorem{expl}{Example}

\newcommand{\ket}[1]{{\left\vert{#1}\right\rangle}}

\begin{document}

\title{Wire Recycling for Quantum Circuit Optimization}	

\author{Alexandru Paler}, 
\affiliation{Johannes Kepler University Linz, Altenberger Str. 69, 4070 Linz, Austria}
\author{Robert Wille},
\affiliation{Johannes Kepler University Linz, Altenberger Str. 69, 4070 Linz, Austria}
\author{Simon J. Devitt}
\affiliation{Center for Emergent Matter Sciences, Riken, Saitama 351-0198, Japan}
\date{\today}
\begin{abstract}
Quantum information processing is expressed using quantum bits (qubits) and quantum gates which are arranged in the terms of quantum circuits. Here, each qubit is associated to a quantum circuit wire which is used to conduct the desired operations. Most of the existing quantum circuits allocate a single quantum circuit wire for each qubit and, hence, introduce a significant overhead. In fact, qubits are usually not needed during the entire computation but only between their initialization and measurement. Before and after that, corresponding wires may be used by other qubits. In this work, we propose a solution which exploits this fact in order to optimize the design of quantum circuits with respect to the required wires. To this end, we introduce a representation of the lifetimes of all qubits which is used to analyze the respective need for wires. Based on this analysis, a method is proposed which ``recycles'' the available wires and, by this, reduces the size of the resulting circuit. Numerical tests based on established reversible and fault-tolerant quantum circuits confirm that the proposed solution reduces the amount of wires by more than 90\% compared to unoptimized quantum circuits. Source code is available at \url{http://github.com/alexandrupaler/wirerecycle}.
\end{abstract}

\maketitle

\section{Introduction}

Quantum computing~\cite{NC00} is an emerging technology that is recently receiving significant attention due to its ability to solve several classes of problems which are computationally inefficient on standard classical computers. Current applications of quantum computing range from integer factorization, unstructured search \cite{Montanaro2016} and quantum chemistry \cite{lanyon2010towards}.

Corresponding solutions are theoretically described in terms of quantum algorithms which, in turn, are realized as \emph{quantum circuits}.
In contrast to classical circuits, a quantum circuit works with so-called \emph{qubits} that cannot only assume the values~1 and~0, but also their superposition.
Here, operations to be conducted are represented in terms of \emph{quantum gates}.
Electronic Design Automation in this domain deals with questions how to synthesize, optimize, and verify corresponding quantum circuits~\cite{svore2006layered}.

Current efforts are mainly motivated by the conclusion that practical quantum circuits need to incorporate a large amount of error-correction techniques because the quantum hardware is highly susceptible to environmental noise~\cite{FMM13}. Practical computations necessitate a high amount of hardware resources, which could be lowered by optimizing quantum circuits~\cite{DSMN13,DMN08}. Accordingly, existing quantum circuit optimization methods focus on either reducing the number of qubits or the number of particular gates, or both~\cite{saeedi2013synthesis}.

However, this view ignores the fact that, eventually, quantum circuits are synthesized so that each single qubit is associated to a \emph{quantum circuit wire}. 
While the qubit is a representation of quantum information, quantum circuit wires represent the structural property of a circuit. 
More precisely, in order to realize a quantum algorithm, a quantum circuit wire is instantiated for each qubit and represents the passage of time. Then, the operations
to be conducted on this qubit are realized in terms of quantum gates applied to the respective wire.
Because of the one-to-one mapping between qubits and wires, the quantum circuit terminology often accepts both terms as being equivalent terms.

But this perspective and the existing design scheme frequently lead to quantum circuits of considerable size with respect to the needed quantum circuit wires.
In fact, qubits are usually not needed during the entire computation of an algorithm, but only during their \emph{lifetime}, i.e.~between their initialization and measurement. 
Before and after that, the quantum circuit wire may be used by other qubits.
But, thus far, existing 
design solutions do not exploit this fact and create a dedicated quantum circuit wire for each qubit -- obviously leading to a significant overhead.

In this work, we propose to ``decouple'' the terms qubit and quantum circuit wire and exploit the potential of \emph{recycling wires} in the design of quantum circuits.
To this end, a method is proposed which associates multiple qubits to the same wire, so that individual qubit lifetimes do not overlap in time. 
While recycling-methods have already been proposed before in~\cite{martin2012experimental} (specific quantum algorithm perspective) and in~\cite{horsman2011reduce} (specific quantum computing architecture perspective), this work exploits recycling for the design of arbitrary quantum circuits.  Similar work from Parent, Roetteler and Svore \cite{REVS}, known as REVS, also examines resource recycling that can be integrated with quantum compilers such as LIQi$|\rangle$.

The proposed methods clearly show that quantum circuit optimization should focus on the number of wires instead of the number of qubits. 
In fact, wire recycling as proposed in this work generates a quantum circuit with the same number of qubits but with less wires and, hence, with a substantially lower hardware requirement. In the best case, the amount of quantum circuit wires can be reduced by more than 90\%. The reduction is proportional to the amount of resources required to implement fault-tolerant quantum circuits, where error correction techniques are heavily used.

The paper is organized as follows: Sec.~\ref{sec:back} will briefly introduce the formalism of quantum circuits. The idea of wire recycling is introduced and illustrated in Sec.~\ref{sec:wirerec}, while Sec.~\ref{sec:method} provides a detailed description of the proposed optimization method. Conclusion and future work are discussed in Sec.~\ref{sec:concl} after having analyzed the performance of the method using relevant reversible and quantum circuits in Sec.~\ref{sec:res}.

\section{Background}
\label{sec:back}

Quantum computations are reversible by definition, meaning that the inputs can be computed from the outputs by applying the circuit backwards. This is possible because 
(1)~quantum circuit gates have an equal number of inputs and outputs and (2)~gates implement a specific type of operation. Classical circuits can be made reversible by using gates implementing bijective Boolean functions, and thus having an equal number of inputs and outputs as well~\cite{NC00}. Classical reversible circuits do not have the computational power of their quantum counterparts, but are described using the same circuit formalism and often serve as initial blueprint~\cite{shende2003synthesis}. 

In the following we consider reversible circuits being a subtype of quantum circuits.

\subsection{Initializations, Gates and Measurements}

\begin{figure}[t]
	\centering
	\includegraphics[width=0.9\columnwidth]{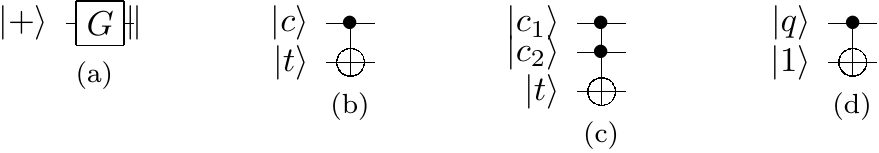}
	\caption{Circuit examples. The $\bullet$ symbol indicates control qubits, and $\oplus$ target qubits: a) a single qubit circuit having the input initialized to $\ket{+}$, operated by a gate $G$ and the qubit is finally measured; b) the CNOT gate; c) the Toffoli gate; d) the qubit initialized to $\ket{1}$ is an input ancilla. Qubit initialization and measurement were left in circuits b-d).}
	\label{fig:circex}
\end{figure}

A quantum circuit is executed by applying a time ordered set of gates, which are generally written from left to right. A time axis running left-to-right can be linked to any quantum circuit, so that each circuit operation can be identified at a certain point in time. Inputs are on the left and outputs on the right. Fig.~\ref{fig:circex} illustrates corresponding examples. A quantum circuit implements three types of operations: 1) qubit initialization; 2) quantum gates which operate on qubits; 3) qubit measurement. The state of a qubit $q$ is represented as the vector $\ket{q}=\alpha_0\ket{0} + \alpha_1\ket{1}$, where $\alpha_0$ and $\alpha_1$ are complex numbers (called amplitudes) that satisfy $|\alpha_0|^2 + |\alpha_1|^2 = 1$. The vectors $\ket{0}$ and $\ket{1}$ are states analogue to the classical bits $0$ and $1$, but a qubit can be in a \emph{superposition} of these two states, where it is both 0 and 1 (e.g. when $\alpha_0 \neq 0$ and $\alpha_1 \neq 0$).

\begin{definition}
A quantum circuit is a sequence of operations applied to a set of qubits.
\end{definition}

Qubits are initialized before any gates can be applied. Commonly used initial states, particularly in the context of error corrected quantum computation are $\ket{0}$, $\ket{1}$, $\ket{+}=\frac{1}{\sqrt{2}}(\ket{0}+\ket{1})$, $\ket{Y}=\frac{1}{\sqrt{2}}(\ket{0}+i\ket{1})$ and $\ket{A}=\frac{1}{\sqrt{2}}\ket{0}+e^{i\pi/4}\ket{1})$.

Quantum gates manipulate the state of qubits, but in contrast to classical gates, quantum gates always have an equal number of inputs and outputs. 
 As a result, quantum gates can operate on single or multiple qubits. Multi-qubit gates are of two types: controlled and uncontrolled. Assuming an $m$-qubit gate named $G$ is applied on the qubit set $Q=\{q_0\ldots q_m\}$. If $G$ would be controlled then the function $O_c$ is applied to the qubit subset $T \subset Q$ (target qubits) depending on the state of the qubit subset $C \subset Q, C \cap T = \emptyset, C \cup T = Q$ (control qubits). If $G$ is uncontrolled, then the function $O_n$ would be applied to all $Q$. The controlled-NOT (CNOT) operation between two qubits $Q=\{c, t\}$, where $C=\{c\}$ and $T=\{t\}$, will perform the NOT operation to $T$ only if the state of the qubits in $C$ is $\ket{1}$, thus $cnot(c,t)=(c, t \oplus c)$, where $\oplus$ is the Boolean XOR function.


Qubit measurement is the quantum counterpart of reading the value of a classical bit. The standard measurement in quantum computation, referred to as a $Z$-basis measurement, measures if the qubit is in the $\ket{0}$ or $\ket{1}$ state, collapsing any superposition inconsistent with the measurement result. Another type of measurement is an $X$-basis measurement, which measures if the qubit is in the $\ket{+}$ state or the $\frac{1}{\sqrt{2}}\left(\ket{0}-\ket{1}\right)$ state.

Quantum gates are chosen from an universal gate set sufficient for approximating any quantum computation with arbitrary precision. The set of all single qubit quantum gates together with CNOT forms an universal gate set. Classical reversible circuits are generally specified using the Toffoli gate, which is a 3-qubit gate applying the NOT operation on the target $t$ if both control qubits $c_1, c_2$ are $\ket{1}$. From a Boolean perspective the Toffoli gate implements the function $toffoli(c_1, c_2, t)=(c_1, c_2, t\oplus c_1c_2)$. However, the Toffoli gate is not quantum computing universal but only classically universal (can be used to implement NAND).


\subsection{Ancilla Qubits}

Quantum circuit qubits can be classified into I/O-qubits and ancillae qubits. The total number of qubits of a circuit is the sum between the number of I/O- and ancillae qubits. If a circuit would be abstracted as a black box, the I/O-qubits are the interface to the box (inputs and outputs) and the ancillae are internal to the box. Circuit inputs and outputs are configurable with respect to the states used for qubit initialization and qubit measurement (e.g. can be $\ket{0}$, $\ket{1}$ or $\ket{+}$), whereas ancillae are initialized and measured into specific states (e.g. always $\ket{A}$). Ancillae are used as temporary workbenches \cite{cuccaro2004new} or for increasing the fault-tolerance of the implemented computation \cite{preskill1998fault}. 

\begin{definition}
An \emph{input ancilla} is a qubit always initialized in the same state, irrespective if its measurement is configurable or not. An \emph{output ancilla} is a qubit always measured in the same state, irrespective if its initialization is configurable or not.
\end{definition}

\subsection{Circuit Wires and Qubit Lifetime}

The temporal ordering of operations introduced by a quantum circuit determines the lifetime of a qubit.

\begin{definition}
The lifetime of a qubit is spanned between its initialization and measurement points in time.
\end{definition}

The lifetime consists of three \emph{stages}: a passive, an active and another passive stage. The length (along the time axis linked to the circuit) of a quantum circuit wire is the abstraction of a qubit's lifetime.

The active stage is the relevant one in a qubit's lifetime and exists between the first and the last gate affecting the qubit. The first passive stage exists between the qubit's initialization and the first gate, while the second passive stage exists between the last gate and the qubit's measurement.

\section{Wire Recycling}
\label{sec:wirerec}

In this work, we propose to recycle quantum circuit wires in order reduce the overhead of quantum circuits. 
The main idea is to compute the temporal ordering of all the relevant qubit lifetimes and, afterwards,
arrange them in a fashion so that the corresponding quantum circuits wires are used and re-used (recycled) as best as possible.

\begin{figure}[h]
	\centering
	\includegraphics[width=0.9\columnwidth]{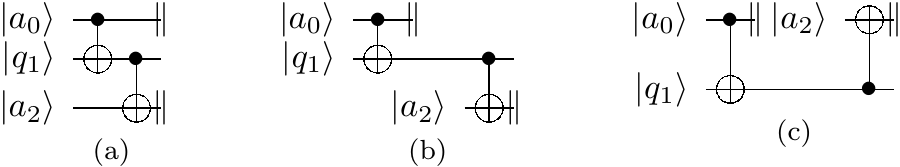}
	\caption{Wire recycling correctness by example: a) a quantum circuit with three wires including two ancillae $a_0$ and $a_2$ and one I/O-qubit $q_1$; b) computation is left unchanged, but the $a_2$ ancilla is initialized after the first CNOT is performed; c) the equivalent computation, but requiring only two wires because both ancillae share the topmost wire. The lifetime of the third qubit $\ket{a_2}$ does not overlap with the lifetime of the first qubit $\ket{a_0}$.}
	\label{fig:wr}
\end{figure}

Note thereby that wire recycling can be applied to ancillae qubits only (only these will be marked with the measurement symbol $\|$), since the lifetime of I/O-qubits is spanned along the entire time axis.
The general idea is illustrated by means of the following example:

\begin{expl}
Fig.~\ref{fig:wr} shows a quantum circuit with eight operations: three qubit initializations (two ancilla, one I/O-qubit), two CNOTs and three qubit measurements (two ancilla, one I/O-qubit with unmarked measurement). The first CNOT gate applied on the qubits $\ket{q_0}$ and $\ket{q_1}$ is applied before the CNOT affecting $\ket{q_1}$ and $\ket{q_2}$, while each of the qubits needs to be firstly initialized and finally measured. The lifetime of qubit $q_0$ starts at the beginning of the first quantum circuit wire, includes the application of the first CNOT where the wire is acting as control and ends at the rightmost wire end point. The lifetime of qubit $q_1$ starts at the beginning of the second wire, includes the application of the first CNOT and the application of the second CNOT. Its lifetime ends at the rightmost end point of the second wire.
Hence, the lifetime  of the third qubit $\ket{a_2}$ does not overlap with the lifetime of the first qubit $\ket{a_0}$ and one quantum circuit wire can be recycled.
This yields the quantum circuit as shown in Fig.~\ref{fig:wr}.
\end{expl}

\section{Method}
\label{sec:method}

In order to introduce wire recycling an equivalent representation of arbitrary quantum circuits is introduced in the form of graphs expressing the temporal relation between circuit operations. The structural analysis of the graphs will be used afterwards to devise two recycling heuristics.

\subsection{Causal Graph}

The temporal ordering of quantum circuit operations can be modeled using the $\rightarrow$ operator by writing $g1 \rightarrow g2$ if gate $g1$ needs to be executed before gate $g2$. The situation exists if the gate $g2$ takes as input one of the outputs of $g1$. Therefore, any qubit initialization will precede a gate, and, for example, we can write $input1 \rightarrow g3$, if the gate $g3$ is applied on the freshly initialized qubit existing at the circuit's input named $input1$. For this reason, qubit measurements will not precede any operation, but will always have their own predecessors. For example, $g4 \rightarrow output2$ indicates that the output qubit of gate $g4$ existing on the circuit's output named $output2$ is measured. The $\rightarrow$ operator is transitive, because if $a \rightarrow b$ and $b \rightarrow c$, then $a \rightarrow c$.

All the temporal relations between quantum circuit operations modeled using the $\rightarrow$ operator can be abstracted by \emph{causal graphs}. A quantum circuit is represented by a graph, where each node has two associated attributes: 1) $type$ (an operation identifier), and 2) $wires$ (the set of wires linked to the qubits operated by the gate). A temporal ordering between two circuit operations is established if one of the operations is based on the output of the other. Thus, if two operations are applied to the same qubit, an edge or a chain of edges will exist between the corresponding operation nodes. Each path in a causal graph represents a chain of circuit operation precedences.

\begin{definition}
A quantum circuit causal graph is a directed acyclic graph where each circuit operation (initialization, gate, measurement) is abstracted by a node, and edge directions indicate the relative temporal ordering of the circuit operations.
\end{definition}

\begin{observation}
A graph edge is abstracting a quantum circuit wire segment existing between the circuit operations represented by the adjacent graph nodes.
\end{observation}


\begin{figure}[t]
\centering
\includegraphics[width=0.9\columnwidth]{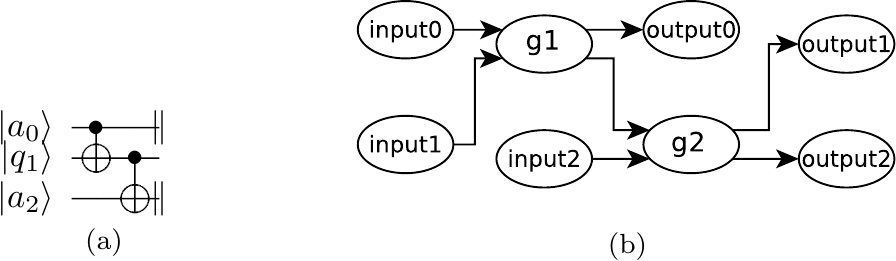}
\caption{a) The circuit from Fig.~\ref{fig:wr}. b) The causal graph of the circuit.}
\label{circ:one}
\end{figure}

An initial causal graph does not contain any direct edges between output and input nodes: there is no precedence established between qubit initializations and  measurements of different qubits. Thus, initially, the precedence of qubit lifetimes is not known.

\begin{expl}
In Fig.~\ref{circ:one}b, the lifetime of $\ket{a_0}$ (top most qubit in Fig.~\ref{circ:one}a) is represented by the subgraph having $input0$ as root and $output0$ as leaf. The lifetime of $\ket{a_2}$ is the subgraph having $input2$ as root and $output2$ as leaf. It can be concluded that the lifetime of $\ket{a_0}$ can precede the lifetime of $\ket{a_2}$, so that the qubits can share the top most wire (Fig.~\ref{fig:wr}).
\end{expl}

\subsection{Recycling Algorithm}

A circuit's causal graph contains all the information necessary for wire recycling where an input graph having each qubit on a distinct wire is gradually transformed to an output graph having multiple qubits on the same wire. We propose two optimization heuristics for the transformation of graphs. Both heuristics are applications of the same transformation rule, which is the result of an observation about the graphs' structure.

\subsubsection{Relation Between Input and Output Nodes}

There are situations when a qubit is measured before another qubit is initialized, and this can be modeled by a supplemental graph edge connecting the corresponding qubit's output node to the other qubit's input node. When adding such an edge it has to be guaranteed that the two nodes are not elements of a direct path. Otherwise, two equivalent problems arise: 1) the graph contains a cycle which is not allowed because causal graphs are acyclic; 2) although the initialization takes place before the measurement, the edge indicates that the measurement precedes the initialization, which is impossible.

\begin{observation}
A direct edge can be drawn between an output and an input node, if no direct path starts at the input and ends at the output node.
\end{observation}

\begin{figure}
\centering
\includegraphics[width=\columnwidth]{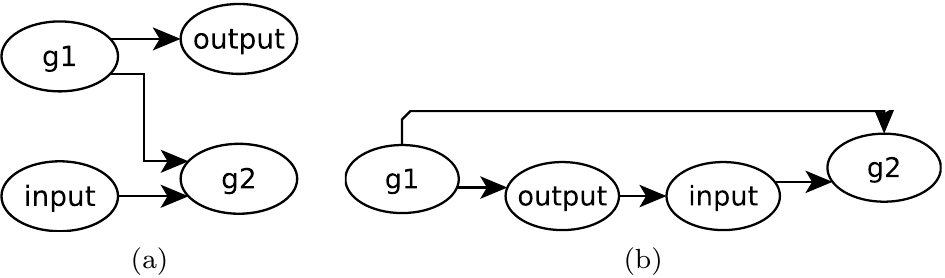}
\caption{The causal graph transformation rule: a) input subgraph; b) output subgraph where a direct edge was added between an output preceding an input.}
\label{circ:red}
\end{figure}

Observation 2 is the basis of the causal graph \emph{transformation rule} (Fig.~\ref{circ:red}) where an edge is added between an output allowed to precede an input (Fig.~\ref{circ:red}a). If the input and the output nodes referenced distinct wires (e.g. $w1$ and $w2$), the input graph can be transformed (Fig.~\ref{circ:red}b), so that both the gates address the same wire, e.g. $w$: 1) gate $g1$ is executed; 2) its output is measured on wire $w$ (node $output$), 3) a new qubit is initialized on wire $w$ (node $input$), 4) gate $g2$ is executed. The result is that $w1$ and $w2$ are treated as segments of $w$ which  abstracts the lifetime of two distinct qubits: the one measured after $g1$, and the one initialized before $g2$. The passive stages of $w1$ and $w2$ were shortened without influencing the computation, the wires do not overlap in time and are arranged so that $w1 \rightarrow w2$.

\subsubsection{Main algorithmic routine}

The central aspect of wire recycling is to determine which qubit measurements are potential candidates to precede qubit initializations. In Alg.~\ref{alg:1} an unoptimized circuit is used to generate a causal graph $cg$ and to infer the nodes corresponding to the input ancillae set ($IA$) and the output ancillae set ($OA$). It is necessary to know how many ancilla output nodes succeed each ancilla input node ($n_a$), because in the light of Observation 2, only outputs preceding inputs can be connected by edges. 

The highest chances of successfully applying the graph transformation rule is for those ancilla inputs used at the latest points in time. Such inputs reach the least outputs and, thus, a maximal number of outputs precedes them. Therefore, for each ancilla input the maximum number of preceding outputs ($|OA| - n_a$) is computed.

The order in which input nodes are connected to outputs nodes is dictated by a priority queue data structure where input nodes are sorted based on the corresponding value of $|OA| - n_a$. For each found output node, the input and all the graph nodes following it will use the same wire as the output.

Because each causal graph node knows the wire(s) it operates on, the notation $a.wire$ is to be interpreted as the wire label of node $a$. In the main algorithmic routine of Alg.~\ref{alg:1}, after an input node $a$ is connected to an output node $o$, the wire label of node $a$ is replaced by the wire label of node $o$, because $a$ will operate on the wire of $o$. 

Finally, after transforming $cg$ by each transformation rule, the queue $Q$ is recomputed to reflect the new graph structure.

\begin{algorithm}
\begin{algorithmic}
\REQUIRE{Input circuit $circ$}
\STATE{$cg =$ causal graph of $circ$}
\STATE{$IA =$ the set of all ancilla input nodes in $cg$}
\STATE{$OA =$ the set of all ancilla output nodes in $cg$}
\STATE{$N = \{n_a | a \in IA, n_a$ the number of nodes $o \in OA$ preceded by $a\}$}
\STATE{$Q =$ queue constructed from $IA$ with elements sorted by priority $|OA| - n_a$}
\WHILE{$Q$ not empty}
	\STATE{$a = $ Read and remove topmost input node from queue $Q$}
	\STATE{$o =$ Result of searching for a preceding output node}
	\IF{Found $o \in OA$}
		\STATE{Draw edge between $o$ and $a$}
		\STATE{Replace at successors of $a$: $a.wire$ with $o.wire$}
	\ENDIF
	\STATE{Update $Q$ to reflect the new structure of $cg$}
\ENDWHILE
\end{algorithmic}
\caption{Qubit Recycling Algorithm}
\label{alg:1}
\end{algorithm}

This work proposes two search heuristics for computing the output node preceding an input node. As previously mentioned, quantum circuit optimization follows circuit synthesis. After synthesis the wires in a circuit diagram can be considered either being ordered or unordered. The first situation is the result of most synthesis methods, as explained in the subsequent example and section. The second situation exists when circuits are further optimized for specific quantum computer architectures (e.g. using nearest neighbor interactions \cite{wille2014exact}).

\begin{figure*}
\includegraphics[width=\textwidth]{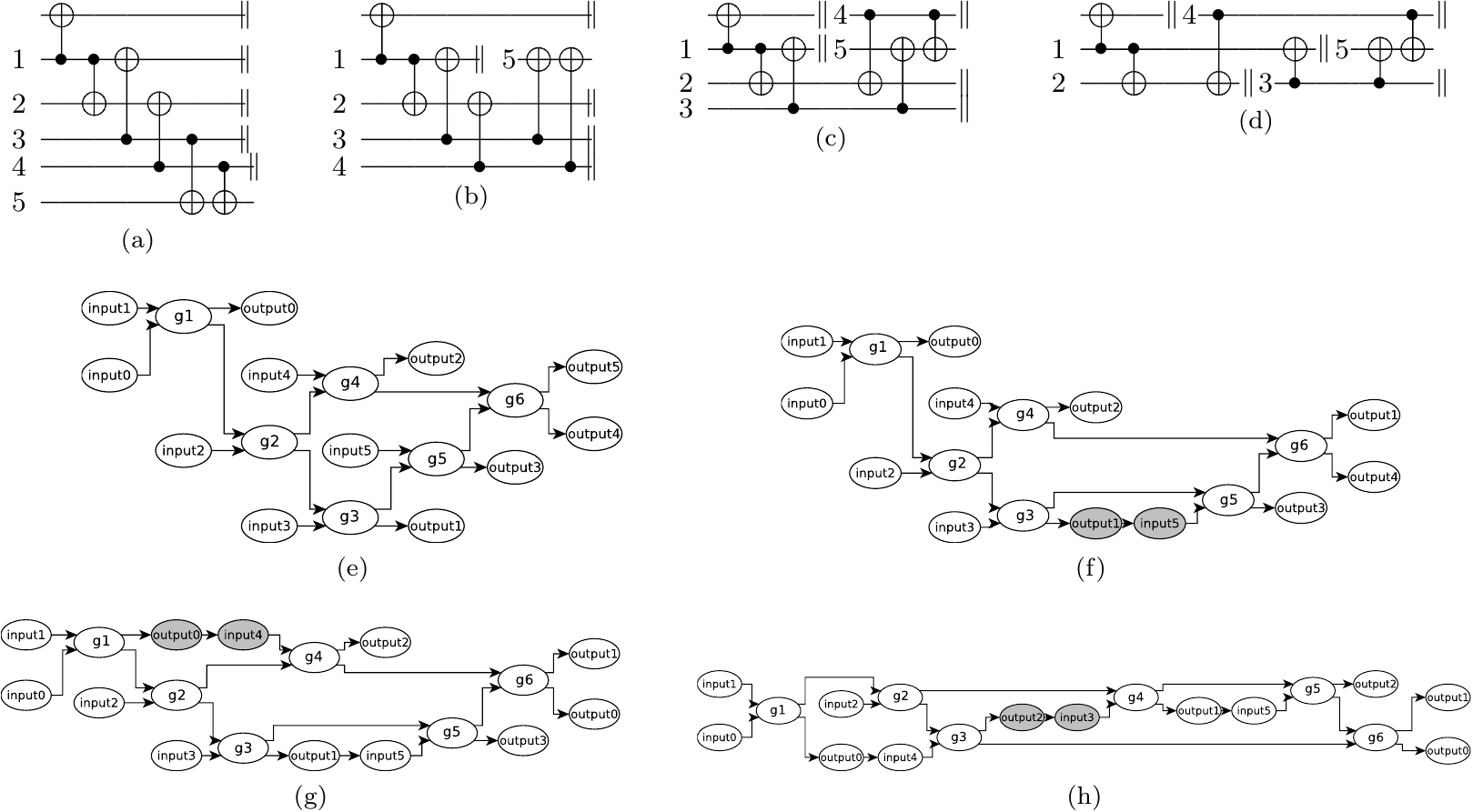}
\caption{Recycling Example (circuit and associated causal graph): a,e) the initial circuit having ancilla inputs labeled and the ancilla outputs marked with $\|$; b,f) $input5 \rightarrow output1$; c,g) $output0 \rightarrow input4$; d,h) $output2 \rightarrow input3$}
\label{fig:example}
\end{figure*}

\begin{expl}
Consider the recycling presented in Fig.~\ref{fig:example}. The wires in Fig.~\ref{fig:example}a are ordered if their labels are considered integers determining their position from the top to the bottom. The wires are unordered if their labels are considered simple strings, implying that the way the circuit was drawn is one of the many possibilities of how its wires can be arranged vertically.
\end{expl}

\subsubsection{M1: Recycling Ordered Wires}

The majority of the synthesis methods introduce ancilla qubits. For a gate sequence $GS$, with gates $g1$ and $g2 \in GS$, $g1 \rightarrow g2$ if gate $g1$ will be executed before $g2$. Gates are processed during synthesis according to their ordering from $GS$ ($g1$ before $g2$), and ancillae are introduced, if necessary, in the same order (the ancilla needed for $g1$ is inserted before the ancilla required by the synthesized $g2$). 

The position where ancillae are introduced into the circuit, e.g. top down numbering of wires in circuit diagram, can be used as an indicator of which wires to recycle. For the same gates, where $g1 \rightarrow g2$, having gate $g1$ operate on the wire numbered $w$ and $g2$ on the one numbered $w+1$, let us assume that the synthesis of $g1$ needs to introduce an ancilla. After its introduction, the ancilla will be numbered $w+1$, and $g2$ will operate on the wire numbered $w+2$. Effectively, the number of wires is increased by one, the ancilla's position is right next to the initial wire of $g1$, and all the wires of the gates succeeding $g1$, e.g. $g2$, are shifted by one position. Therefore, the vertical position of such ancillae relative to the neighboring wires can be used as an indicator of the starting time point of the corresponding qubit's active stage.

This implies that, there exists an ordering between the wires of a synthesized circuit: ancillae are positioned next to the wires representing the specified gates. This observation can be translated to the causal graphs abstracting circuit diagrams: wire ordering is an indicator of the closest output node in time which precedes a currently investigated input node. The first search heuristic is to simply search for the minimum of $o.wire - a.wire$, for the output ancilla $o \in OA$ and the input ancilla $a \in IA$ where $o \rightarrow a$.

\subsubsection{M2: Recycling Unordered Wires}

The previous ordering of wires is not always guaranteed, because a synthesized circuit could have been optimized after synthesis in ways which affected the ordering of wires. As a consequence, wires should not always be assumed ordered.

An unknown wire ordering increases the complexity of the preceding output node search. The heuristic  proposed in this section allows edges to be traversed backwards (Alg.~\ref{alg:2}). Although, causal graphs are directed, some of the edges will need to be traversed backwards due to the structure of the transformation rule (Fig.~\ref{circ:red}a,~\ref{circ:red}b): before recycling there is no direct path connecting the affected input and output nodes. Thus, in order to connect two arbitrary nodes which satisfy the transformation rule, a graph search has to be performed, where at least one edge has to be traversed backwards. 

An aspect not mentioned previously is that, without loss of correctness, the edges in the transformation rule from Fig.~\ref{circ:red}a represent not only a single graph edge but chains of directed edges. Therefore, the output node search is allowed to traverse multiple edges in their specified (forward) direction or backwards. The number of edges traversed backwards ($nr$) is an indication of how much earlier on the circuit's time axis the currently investigated node exists relative to the input node. The selection criteria chosen in Alg.~\ref{alg:2} is to consider the ancilla output node assumed to be closest in time to the input node: forward traversals are preferred to backward traversals. Overall, the shortest path with the lowest $nr$ between an input and and output node is chosen as a solution.

\begin{algorithm}
\begin{algorithmic}
\REQUIRE{Causal graph $cg$; Input node $input$; Current node $current$; Number of edges traversed backwards $nr$}
\STATE{Return $current$ if $\in OA$ and $nr \neq 0$}
\FORALL{$node \in cg, current \rightarrow node$}
	\RETURN{Alg.~\ref{alg:2} with parameters $cg,input,node,nr$}
\ENDFOR
\FORALL{$node \in cg, node \rightarrow current$}
	\RETURN{Alg.~\ref{alg:2} with parameters $cg,input,node,nr+1$}
\ENDFOR
\end{algorithmic}
\caption{Search Preceding Output in a Quantum Circuit with Unordered Wires}
\label{alg:2}
\end{algorithm}

\section{Results}
\label{sec:res}

Wire recycling was prototypically implemented and applied to quantum as well as reversible circuits defined over arbitrary gate sets. This is possible because the graph nodes represent operations which are independent of the implemented computation. The applicability of the proposed method is investigated using reversible circuits and quantum circuits consisting entirely of qubit initializations, CNOTs and qubit measurements (ICM circuits) \cite{paler2015fully}. The source code is available at \url{http://github.com/alexandrupaler/wirerecycle}.

\begin{table*}[t!]
\caption{Reversible circuit optimization results (heuristics M1 and M2) and percentage of recycled wires (\%M1 and \%M2).}
\label{tbl:rev}
\centering
\tiny
\begin{tabular}{lrrrrrr || lrrrrrr}
Circuit & Qubits & Ancilla & M1 & M2 & \%M1 & \%M2 & Circuit & Qubits & Ancilla & M1 & M2 & \%M1 & \%M2\\
\hline
pdc\_307 & 619 & 603 & 464 & 366 & 75 & 59 & sys6-v0\_144 & 10 & 4 & 3 & 2 & 30 & 20\\
spla\_315 & 489 & 473 & 401 & 276 & 82 & 56 & 4mod5-bdd\_287 & 7 & 3 & 2 & 2 & 29 & 29\\
lu\_326 & 299 & 233 & 194 & 185 & 65 & 62 & squar5\_261 & 13 & 8 & 2 & 2 & 15 & 15\\
e64-bdd\_295 & 195 & 130 & 114 & 86 & 58 & 44 & rd53\_138 & 8 & 3 & 2 & 1 & 25 & 13\\
ex5p\_296 & 206 & 198 & 107 & 87 & 52 & 42 & rd53\_139 & 8 & 3 & 2 & 1 & 25 & 13\\
hwb9\_304 & 170 & 161 & 76 & 81 & 45 & 48 & rd73\_140 & 10 & 3 & 2 & 1 & 20 & 10\\
lu\_327 & 203 & 137 & 71 & 36 & 35 & 18 & rd73\_141 & 10 & 3 & 2 & 1 & 20 & 10\\
add64\_184 & 193 & 64 & 63 & 63 & 33 & 33 & sqr6\_259 & 18 & 12 & 2 & 1 & 11 & 6\\
add64\_186 & 193 & 64 & 63 & 63 & 33 & 33 & sym9\_146 & 12 & 3 & 2 & 1 & 17 & 8\\
hwb8\_303 & 112 & 104 & 52 & 52 & 46 & 46 & sym9\_192 & 12 & 3 & 2 & 1 & 17 & 8\\
bw\_291 & 87 & 82 & 44 & 32 & 51 & 37 & wim\_266 & 11 & 7 & 2 & 1 & 18 & 9\\
add32\_183 & 97 & 32 & 31 & 31 & 32 & 32 & dc2\_222 & 15 & 7 & 2 & 0 & 13 & 0\\
add32\_185 & 97 & 32 & 31 & 31 & 32 & 32 & cmb\_214 & 20 & 4 & 1 & 3 & 5 & 15\\
hwb7\_302 & 73 & 66 & 30 & 31 & 41 & 42 & 5xp1\_194 & 17 & 10 & 1 & 2 & 6 & 12\\
cycle10\_293 & 39 & 27 & 20 & 19 & 51 & 49 & inc\_237 & 16 & 9 & 1 & 2 & 6 & 13\\
hwb6\_301 & 46 & 40 & 20 & 15 & 43 & 33 & 4mod5-v0\_19 & 5 & 1 & 1 & 1 & 20 & 20\\
rd84\_313 & 34 & 26 & 20 & 15 & 59 & 44 & 4mod5-v0\_20 & 5 & 1 & 1 & 1 & 20 & 20\\
ham15\_298 & 45 & 30 & 19 & 16 & 42 & 36 & 4mod5-v1\_22 & 5 & 1 & 1 & 1 & 20 & 20\\
add16\_174 & 49 & 16 & 15 & 15 & 31 & 31 & 4mod5-v1\_24 & 5 & 1 & 1 & 1 & 20 & 20\\
add16\_175 & 49 & 16 & 15 & 15 & 31 & 31 & 4mod5-v1\_25 & 5 & 1 & 1 & 1 & 20 & 20\\
sym9\_317 & 27 & 18 & 15 & 11 & 56 & 41 & add6\_196 & 19 & 7 & 1 & 1 & 5 & 5\\
mod5adder\_306 & 32 & 26 & 13 & 11 & 41 & 34 & alu2\_199 & 16 & 6 & 1 & 1 & 6 & 6\\
rd73\_312 & 25 & 18 & 13 & 10 & 52 & 40 & alu3\_200 & 18 & 8 & 1 & 1 & 6 & 6\\
plus127mod8192\_308 & 25 & 12 & 10 & 10 & 40 & 40 & alu-bdd\_288 & 7 & 2 & 1 & 1 & 14 & 14\\
plus63mod8192\_310 & 25 & 12 & 10 & 10 & 40 & 40 & apla\_203 & 22 & 12 & 1 & 1 & 5 & 5\\
hwb5\_300 & 28 & 23 & 9 & 9 & 32 & 32 & cm42a\_207 & 14 & 10 & 1 & 1 & 7 & 7\\
plus63mod4096\_309 & 23 & 11 & 9 & 9 & 39 & 39 & cm85a\_209 & 14 & 3 & 1 & 1 & 7 & 7\\
cm163a\_213 & 29 & 13 & 9 & 6 & 31 & 21 & decod24-enable\_124 & 6 & 3 & 1 & 1 & 17 & 17\\
add8\_172 & 25 & 8 & 7 & 7 & 28 & 28 & decod24-enable\_125 & 6 & 3 & 1 & 1 & 17 & 17\\
ham7\_299 & 21 & 14 & 7 & 7 & 33 & 33 & example2\_231 & 16 & 6 & 1 & 1 & 6 & 6\\
cu\_219 & 25 & 11 & 6 & 6 & 24 & 24 & f2\_232 & 8 & 4 & 1 & 1 & 13 & 13\\
rd84\_142 & 15 & 7 & 6 & 4 & 40 & 27 & f51m\_233 & 22 & 8 & 1 & 1 & 5 & 5\\
rd84\_143 & 15 & 7 & 6 & 4 & 40 & 27 & frg1\_234 & 31 & 3 & 1 & 1 & 3 & 3\\
alu1\_198 & 20 & 8 & 5 & 6 & 25 & 30 & misex1\_241 & 15 & 7 & 1 & 1 & 7 & 7\\
dk27\_225 & 18 & 9 & 5 & 6 & 28 & 33 & pm1\_249 & 14 & 10 & 1 & 1 & 7 & 7\\
sym6\_316 & 14 & 8 & 5 & 4 & 36 & 29 & radd\_250 & 13 & 5 & 1 & 1 & 8 & 8\\
rd53\_311 & 13 & 8 & 5 & 2 & 38 & 15 & rd32\_270 & 5 & 2 & 1 & 0 & 20 & 0\\
dk17\_224 & 21 & 11 & 3 & 3 & 14 & 14 & rd32\_271 & 5 & 2 & 1 & 0 & 20 & 0\\
mini\_alu\_305 & 10 & 6 & 3 & 3 & 30 & 30 & 4gt11\_83 & 5 & 1 & 0 & 1 & 0 & 20\\
pcler8\_248 & 21 & 5 & 3 & 3 & 14 & 14 & 4gt5\_75 & 5 & 1 & 0 & 1 & 0 & 20\\
x2\_267 & 17 & 7 & 3 & 3 & 18 & 18 & 4gt5\_77 & 5 & 1 & 0 & 1 & 0 & 20\\
sys6-v0\_111 & 10 & 4 & 3 & 2 & 30 & 20 & mlp4\_245 & 16 & 8 & 0 & 1 & 0 & 6
\end{tabular}
\end{table*}

\subsection{Reversible Circuits}
The entire set of RevLib \cite{wille2008revlib} circuits was downloaded and wire recycling was applied to each circuit using the methods M1 and M2. The results are reported in Table~\ref{tbl:rev}. The columns $\%M1$ and $\%M2$ are computed as the ratios $\frac{M1}{Qubits}$ respectively $\frac{M2}{Qubits}$. It can be noticed that the highest optimization rate is about $80\%$ for $M1$, and that in general $M1$ performs better than $M2$. This can be explained by the fact that M1 uses the knowledge about how synthesis methods work: there is a very high probability that ancillae are introduced in an ordered fashion, which greatly benefits M1. The large reduction range ($0\% \ldots 80\%$) illustrates that the best optimizations are achieved for circuits having qubits with short lifetimes and interacted by few gates.

\subsection{ICM Circuits}

Wire recycling was evaluated on fault-tolerant implementations of the quantum adder circuits presented in \cite{cuccaro2004new}. The fault-tolerant form was obtained after transforming the circuits using the method from \cite{paler2015fully}. Fault-tolerant quantum circuits include only qubit (I)nitializations, (C)NOT gates and qubit (M)easurements. ICM circuits use only CNOT gates, because the necessary single qubit gates are implemented by teleportation mechanisms \cite{NC00,paler2015fully} with ancilla qubits initialized in $\ket{0}$, $\ket{+}$, $\ket{Y}$ or $\ket{A}$. Qubits can be be measured in the $X,Z$ basis \cite{paler2015fully}.

\begin{table}
\caption{Quantum adder circuit optimization results and percentage of recycled wires (\%M1 and \%M2).}
\label{tbl:cucc}
\centering
\tiny
\begin{tabular}{lrrrrrr}
Circuit & Qubits & Ancilla & M1 & M2 & \%M1 & \%M2\\
\hline
Cuccaro4 & 304 & 295 & 276 & 251 & 91\% & 83\% \\
Cuccaro5 & 390 & 379 & 357 & 311 & 92\% & 80\% \\
Cuccaro6 & 476 & 463 & 438 & 382 & 92\% & 80\% \\
Cuccaro7 & 562 & 547 & 520 & 451 & 93\% & 80\% \\
Cuccaro8 & 648 & 631 & 601 & 531 & 93\% & 82\% \\
Cuccaro9 & 734 & 715 & 681 & 592 & 93\% & 81\% \\
Cuccaro10 & 820 & 799 & 761 & 669 & 93\% & 82\% \\
Cuccaro11 & 906 & 883 & 841 & 724 & 93\% & 80\% \\
Cuccaro12 & 992 & 967 & 920 & 798 & 93\% & 80\% \\
Cuccaro13 & 1078 & 1051 & 1000 & 884 & 93\% & 82\% \\
Cuccaro14 & 1164 & 1135 & 1080 & 949 & 93\% & 82\% \\
Cuccaro15 & 1250 & 1219 & 1159 & 1009 & 93\% & 81\% \\
Cuccaro16 & 1336 & 1303 & 1238 & 1083 & 93\% & 81\% \\
Cuccaro17 & 1422 & 1387 & 1318 & 1164 & 93\% & 82\% \\
Cuccaro18 & 1508 & 1471 & 1397 & 1231 & 93\% & 82\% \\
Cuccaro19 & 1594 & 1555 & 1477 & 1279 & 93\% & 80\% \\
Cuccaro20 & 1680 & 1639 & 1557 & 1363 & 93\% & 81\%
\end{tabular}
\end{table}

The results reported in Table~\ref{tbl:cucc} show that in general M1 performs better than M2. At the same time, more than $90\%$ wire reduction obtained for all the circuits shows that fault-tolerant circuit synthesis introduces a vast amount of ancillae with very short active stages.

\section{Conclusion}
\label{sec:concl}

This work introduced quantum circuit wire recycling as a method for quantum circuit optimization. To this end causal graphs were introduced as an equivalent description of a quantum circuit and a recycling method was formulated using two heuristics applied to causal graphs. The method was implemented and its performance was evaluated on circuits from the RevLib circuit library and fault-tolerant implementations of quantum adder circuits. Optimization rates of more than  $90\%$ showcase the potential of wire recycling to reduce the quantum hardware requirements. Future work will analyze the complexity of the method, the temporal overhead introduced by recycling, and ways to optimize it using classic data flow graph algorithms.

\section{Acknowledgments}
SJD acknowledge support from the JSPS grant for challenging exploratory research and the JST ImPACT project. This work has partially been supported by the European Union through the COST Action IC1405. 
\label{sec:concl}


\bibliographystyle{unsrt}
\bibliography{recycle} 

\end{document}